\documentclass[aps,twocolumn,showpacs]{revtex4}
\usepackage{graphicx}

\begin{document}

\title{A Bose-Einstein condensate in a random potential}

\author{J.~E.~Lye$^{*}$, L.~Fallani, M.~Modugno$^{1}$, D.~Wiersma, C.~Fort, and M.~Inguscio}

\affiliation{LENS, Dipartimento di Fisica, and INFM Universit\`a
di Firenze via Nello
Carrara 1, I-50019 Sesto Fiorentino (FI), Italy\\
$^{1}$ also BEC-INFM Center, Universit\`a di Trento, I-38050 Povo
(TN), Italy}

\begin{abstract}
An optical speckle potential is used to investigate the static and
dynamic properties of a Bose-Einstein condensate in the presence
of disorder. For strong disorder the condensate is localized in
the deep wells of the potential. With smaller levels of disorder,
stripes are observed in the expanded density profile and strong
damping of dipole and quadrupole oscillations is seen.
Uncorrelated frequency shifts of the two modes are measured for a
weak disorder and are explained using a sum-rules approach and by
the numerical solution of the Gross-Pitaevskii equation.
\end{abstract}



\maketitle

The nature of a bosonic system in the presence of disorder has
been intensely explored both theoretically and experimentally,
particularly in the context of $^{4}$He in a porous material,
where the breakdown of superfluidity with sufficient disorder was
observed \cite{review}. In the zero temperature limit, theoretical
studies have provided the striking result that there exists a
regime where the superfluid fraction is significantly smaller than
the condensate fraction \cite{lesssuperfluid}. A further
consequence of disorder is the quantum phase transition from a
superfluid to a localized Bose-Glass state \cite{boseglass}.
Indications of such a state have been found in several physical
systems but the precise phase diagram of the progression from
superfluid, to Bose-Glass, to Mott-insulator state remains unclear
\cite{fromdamski}. Furthermore, Anderson localization occurs for
weakly interacting bosons \cite{anderson58}, and plays an
important role in solid state physics where electron transport can
be disrupted by defects in a solid \cite{soliddefects}, and more
recently was observed for photons in strongly scattering
semiconductor powders \cite{Diederik97}.

The realization of Bose-Einstein condensation (BEC) in a dilute
alkali vapor offers the possibility of an ideal system in which to
explore the rich arena of disorder related phenomena. Already
there has been considerable work on the superfluidity and
long-range coherence properties of these degenerate Bose gases,
and with the combination of BEC and optical lattices there has
been an explosion of observations on phenomena of solid
state-physics \cite{latticereview}, notably including the
superfluid to Mott-Insulator quantum phase transition
\cite{greiner}. Recently, theoretical works have discussed the
possibility of Bose-Glass and Anderson localization transition in
a BEC in a disordered optical lattice \cite{RothandBurnett,
Damski}. Understanding the effect of disorder is also important
for BEC in microtraps where fragmentation of the trapped BEC
density profile has been observed and is now attributed to
intrinsic disorder in the fabrication of the microchip
\cite{fragmentation, Fortagh}. It is crucial for integrated atom
optics on microchips to quantitatively define how this disorder
will change the coherence and transport properties, particularly
considering that in 1D disordered systems localization of the
excitations could play an important role \cite{shlyapnikov}.

\begin{figure}[h!]
\begin{center}
\includegraphics[width=0.8\columnwidth]{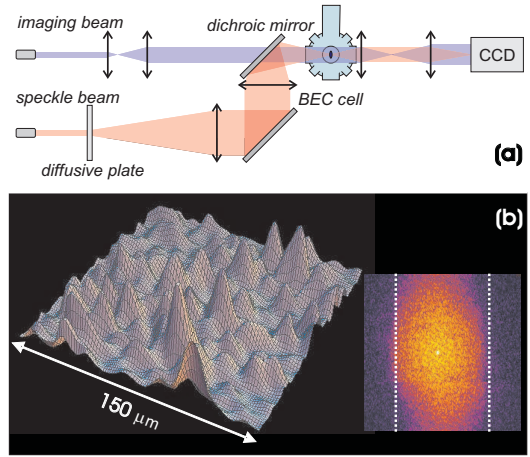}
\end{center}
\caption{(a) Optical setup for both the speckle potential and the
imaging beam for the BEC. The axial direction of the magnetic trap
is in the vertical direction of the figure. (b) 3D representation
of the speckle potential (left) and its Fourier transform (right).
The dotted lines correspond to a length scale of about 10~$\mu$m
in the axial direction.} \label{setup}
\end{figure}
We present in this Letter the first results on a Rb BEC in a
disordered potential, where the precise realization of the
disorder is controllable. We consider both the limit of strong
disorder, where we observe that the BEC is localized in the wells
of the high random potential, and the situation where the disorder
is smaller than the chemical potential of the BEC, where we see
stripes in the expanded density profile and damping of dipole
oscillations. In the perturbative limit of very weak disorder we
observe shifts in the frequency of the quadrupole excitation mode.

To produce the random potential a laser beam is shone through a
diffusive plate. The beam is derived from a Ti:Sa laser at
$\lambda=822$~nm, far detuned with respect to the Rb D1 line at
$\lambda=795$~nm. The resulting speckle pattern is imaged onto the
trapped BEC, as shown in Fig.~\ref{setup}(a), producing a
two-dimensional (2D) potential perpendicular to the beam
propagation, that varies spatially but is stable in time. A nice
feature of the optical setup is that the path of the speckle beam
is combined with the path of the imaging beam using a dichroic
mirror ($\lambda>800$~nm reflected, $\lambda<800$~nm transmitted).
The imaging setup is used to detect both the BEC and the speckle
pattern, and thus we can image in consecutive photos the position
of the trapped BEC and the precise realization of the random
potential that the BEC experiences. The BEC is produced in a
Ioffe-Pritchard magnetic trap, elongated perpendicular with
respect to the speckle beam. The trapping frequencies are
$\omega_z=2\pi \times (8.74\pm0.03)$~Hz axially and
$\omega_\perp=2\pi \times (85\pm1)$~Hz radially, with the axis of
the trap oriented horizontally. Our typical BECs are made of
$\simeq 3 \times 10^5$ atoms in the hyperfine ground state
$|F=1;m_F=-1>$, with a peak density $n\simeq1.2 \times
10^{14}$~cm$^{-3}$.

A 3D representation of a typical speckle potential is displayed in
Fig.~\ref{setup}(b). In the right part Fig.~\ref{setup}, we show
the Fourier transform of this potential, where the smallest length
scale of the speckle is 10~$\mu$m. The average distance between
neighboring speckles is approximately 20~$\mu$m. The axial and
radial size of our BEC is 110~$\mu$m and 11~$\mu$m respectively.
The BEC probes 6 wells in the axial direction and only 1 well on
average in the radial direction, and thus the cylindrical geometry
enforces a quasi-one dimensionality on the speckle-condensate
system. We define the speckle height $V_s$ by taking twice the
standard deviation of the speckle potential over a distance of
200~$\mu$m along the BEC axial direction.  The speckle height is
expressed throughout this Letter in units of frequency.

\begin{figure}[h!]
\begin{center}
\includegraphics[width=0.8\columnwidth]{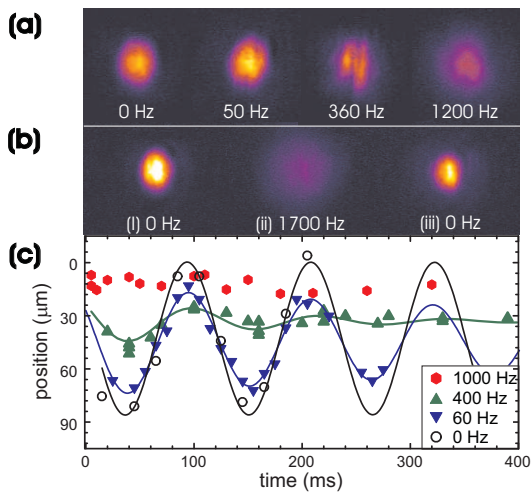}
\end{center}
\caption{(a) Density profile of the BEC after 28~ms of expansion
from the combined magnetic and speckle potential for varying
speckle potential intensities $V_s$ (indicated in bottom part). To
verify that the transfer to the speckle is adiabatic (b) shows the
density profiles with (i) no speckle potential, (ii) when the
speckle potential is abruptly switched off, and (iii) when it is
adiabatically ramped off. (c) Dipole oscillations in the combined
magnetic and speckle potential for varying $V_s$.}
\label{rateszoom}
\end{figure}

An initial insight into the effect of the random potential can be
obtained by considering the BEC in expansion.
Fig.~\ref{rateszoom}(a) shows a series of absorption images after
28~ms of free expansion for different values of $V_s$. In each
image the speckle intensity is increased adiabatically using a
200~ms exponential ramp with a time constant of 40~ms. After 50~ms
in the combined magnetic and speckle potential, both potentials
are abruptly switched off. For $V_s\sim$50~Hz much lower than the
chemical potential of the BEC in the harmonic trap ($\sim$1~kHz),
no change from the standard Thomas-Fermi (TF) profile is seen. As
the potential is increased to an intermediate power ($\sim$300~Hz)
we see a strong modification in the expanded density profile. Such
modifications can be a signature of the development of different
phase domains across the BEC, similar to the stripes seen with
highly elongated quasi-condensates \cite{Sengstock, Richard03}, or
when different momentum components are present due to the growth
of instabilities \cite{DI}. We note that the spacing of the
stripes we observe, $>50$~$\mu$m, can not be related to the
characteristic length scale of the speckle pattern, 10-20~$\mu$m,
the latter would give rise to an interferogram length scale of
6-12~$\mu$m after 28~ms of expansion. Finally, when $V_s>$1~kHz, a
value greater than the chemical potential, the system enters the
tight-binding regime. Here in expansion only a broad gaussian
profile is seen. The tunnelling between wells is expected to be
completely suppressed for such large spacing of adjacent sites
($>10$~$\mu$m), and we are seeing the expansion of an array of
randomly spaced sources. Unlike the situation of the expansion of
a 1D array of equally spaced sources with random phases where a
periodic interference pattern is still visible in a single run
\cite{Dalibard}, no regular interference is expected to be seen
with random spacing either for uniform or for random phases. To
check that the ramp is adiabatic, we have ramped the speckle on
(to our maximum speckle height of 1.7~kHz) and then off, using the
exponential ramp described earlier. Fig.~\ref{rateszoom}(b)(i)
shows the expanded BEC density profile with no speckle, (ii) the
expanded profile when the speckle potential is abruptly switched
off, and (iii) the recovered TF profile when the speckle intensity
is adiabatically decreased. This also clearly indicates that the
density profile in Fig.~\ref{rateszoom}(b)(ii) can not be
attributed to heating of the atomic cloud.

The transport properties in each of the above regimes are
investigated by abruptly shifting the magnetic trap in the axial
direction by 25~$\mu$m inducing dipole oscillations shown in
Fig.~\ref{rateszoom}(c). Undamped oscillations are observed in the
absence of the speckle. The dipole motion for very weak $V_s$
($\sim$60~Hz) is slightly damped with an unchanged frequency, and
in the intermediate 400~Hz potential, where phase fragmentation is
seen in expansion, strong damping is observed. Once in the
tight-binding regime, $V_s>$1~kHz, the atomic cloud does not
oscillate and remains localized in the deep, largely spaced, wells
of the speckle potential on the side of the magnetic trap. This is
analogous to the `pinning' effect seen in magnetic microtraps,
where the axial frequency is relaxed but the atoms do not
propagate along the newly created waveguide remaining localized in
the local minima due to the disorder in the microchip
\cite{Fortagh}.

\begin{figure}[t!]
\begin{center}
\includegraphics[width=0.95\columnwidth]{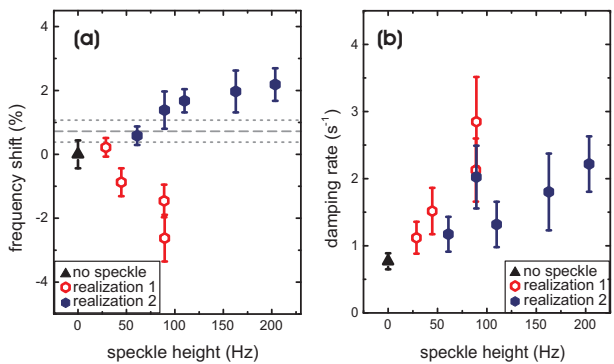}
\end{center}
\caption{Quadrupole oscillations with increasing $V_s$ for two
different speckle realizations: (a) frequency shift (relative to
the quadrupole frequency $13.72\pm 0.06$~Hz measured in the pure
harmonic trap), and (b) damping rate. The dashed line corresponds
to $\sqrt{5/2}$~$\nu_{d}$, calculated from the measured dipole
frequency of $8.74\pm0.03$~Hz. The error (dotted lines) reflects
the variation of the trap over several months.} \label{fig3}
\end{figure}
Precise information on the nature of BEC has been gained through
studies of collective excitations \cite{stringarireview} and it is
therefore a natural starting point to gather information on a BEC
in a random potential. We measure the frequency of both the
dipole, $\nu_d$, and quadrupole, $\nu_q$, modes of a BEC confined
in the harmonic magnetic trap combined with a weak speckle
potential in the range of 30 - 200~Hz corresponding to a small
perturbation of the harmonic potential. The quadrupole mode is
excited with a resonant modulation of the radial trapping
frequency at 13.7~Hz, producing oscillations of 5$\%$ of the
trapped BEC axial width, while the dipole mode is excited by an
abrupt shift of the trap of 25~$\mu$m in the axial direction. In
both cases the speckle potential remains stationary. The results
for the frequency shift of the quadrupole oscillations versus
$V_s$ for two different speckle realizations are shown in
Fig.~\ref{fig3}, with the corresponding damping of this mode. The
shift is taken relative to the measured quadrupole frequency,
$13.72\pm 0.06$~Hz, in the pure harmonic trap. We see frequency
shifts of up to 3$\%$, and also observe that both the sign and
amplitude of the shift depend on the exact realization of the
speckle potential. We measure no change in the dipole frequency
within our experimental resolution of $1\%$. These measurements
are in contrast to what is seen with a regular standing-wave
optical lattice, where both $\nu_d$ and $\nu_q$ are rescaled with
the single particle effective mass due to the optical lattice
\cite{chiara}.

In the absence of the speckle potential, the quadrupole mode of an
elongated BEC is given by $\nu_{q}=\sqrt{5/2}\nu_d$ in the
zero-temperature TF approximation \cite{stringari96}. In the
presence of an additional potential, this value can be modified by
both the deviation of the potential from a pure harmonic
oscillator, and other effects such as changes in the condensate
and superfluid fraction, modification of the interatomic
interactions, or an increase in the thermal component. For
example, in the collisional hydrodynamic limit, calculated for a
gas above the condensation temperature, the frequency of the
quadrupole mode in a harmonic potential is
$\nu_{q}=\sqrt{12/5}\nu_d$, while in the non-interacting thermal
limit $\nu_{q}=2\nu_d$ \cite{griffen97}. Actually, even without
the speckle potential, our measured quadrupole frequency is
shifted $-0.7\%$ with respect to the predicted $\sqrt{5/2}\nu_d$,
similar to what has been measured previously and attributed to the
presence of a residual thermal component \cite{Stamper-Kurn98}. To
check that the observed frequency shifts with the speckle present
are not simply an effect of heating, we have measured the
quadrupole frequency in the presence of a large thermal fraction
(50$\%$) observing only a small shift of -0.4$\%$ with respect to
the measurement with no discernable thermal fraction.

To understand the effect of a shallow random potential on the
dynamics of the BEC we use two theoretical approaches. The
Gross-Pitaevskii equation (GPE) is solved in the actual potential
used in the experiment, and the numerical predictions are
confirmed using a sum-rules approach.
\begin{figure}[t!]
\begin{center}
\includegraphics[width=0.95\columnwidth]{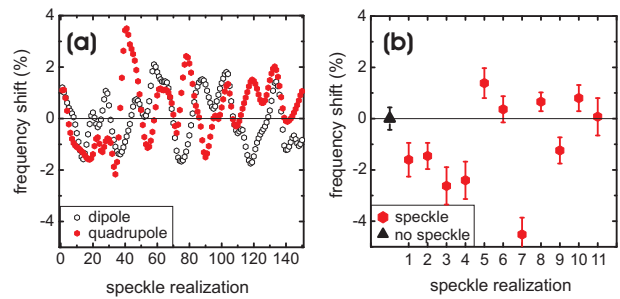}
\end{center}
\caption{(a) Dipole and quadrupole frequency shifts for 150 slices
of the same speckle realization calculated from the sum rules
approach. (b) Measured shift in the quadrupole frequency for 11
different realizations of the speckle at a fixed height of
$80\pm10$~Hz. } \label{fig4}
\end{figure}
We investigate the dynamics of the BEC with the GPE in the
combined speckle, $V_{opt}$, and harmonic, $V_{ho}$, potential
\begin{eqnarray}
i\hbar \frac{\partial \Psi}{\partial t} & = & \left(
-\frac{\hbar^2}{2m}\nabla^2 + V_{opt} + V_{ho} + \frac{4\pi
\hbar^2 a}{m}|\Psi|^2 \right) \Psi \label {GPE}
\end{eqnarray}
where $\Psi$ is the complex BEC order parameter, $m$ is the atomic
mass, and $a$ is the scattering length. The GPE is solved after a
sudden displacement $\Delta z$ of the harmonic trap for the dipole
mode or after a sudden change in the axial trapping frequency for
the quadrupole mode, for a variety of different speckle
realizations. In the case of small amplitude oscillations ($\Delta
z = 3 \mu$m and quadrupole oscillations of $5\%$ of the BEC aspect
ratio), the BEC oscillates coherently with no appreciable damping
for simulations up to 600~ms. The frequencies are red or blue
shifted depending on the particular speckle realization, and in
some configurations a red shift is observed for one mode and a
blue shift for the other. The result for the quadrupole mode shows
an increasing frequency shift with increasing $V_s$ as
qualitatively confirmed by the experimental results in
Fig.~\ref{fig3}a. We found that the predicted frequency shift of
the dipole mode reduces for larger amplitude oscillations since in
this case the BEC probes more of the outer unperturbed harmonic
oscillator potential, and less of the central perturbed potential.
Significant frequency shifts are observed ($>1\%$) only for small
amplitude oscillations ($\Delta z < 10 \mu$m), but unfortunately
we do not have sufficient experimental accuracy with the small
amplitude oscillations to measure these shifts. The somewhat
surprising behavior predicted by the GPE of disparate dipole and
quadrupole frequency shifts can be explained by using the sum
rules approach  and considering $V_{opt}$ as a perturbation.
According to the sum rules, the frequencies of the low lying
collective excitations can be estimated as
$\hbar^2\omega^2={m_3}/{m_1}$ where the moments $m_i$ can be
expressed as commutators involving the many-body hamiltonian $H$
and a suitable excitation operator $F$ \cite{stringarireview}. In
our case this operator can be chosen as $F_d \sim z$ and $F_q \sim
r^2 -\alpha z^2$ for the dipole and quadrupole modes respectively,
$\alpha$ being a variational parameter \cite{kimura}. Treating the
speckle potential $V_{opt}$ as a small perturbation, and writing
$\omega^2=\omega_0^2+\Delta$ we get
\begin{eqnarray}
\Delta_d&\simeq&\frac{1}{m}\langle\partial_z^2V_{opt}\rangle_0
\\
\Delta_q&\simeq&\frac{1}{m}\frac{\langle
z\partial_zV_{opt}+z^2\partial_z^2V_{opt}\rangle_0}{\langle
z^2\rangle_0}
\end{eqnarray}
where the second line is obtained assuming a strongly elongated
BEC. These analytical results demonstrate how the dipole and
quadrupole frequencies have different dependencies on the
particular characteristics of the perturbing potential. We have
verified that these predictions are in agreement with the full
solution of the GPE in the cases considered.

In Fig.~\ref{fig4}(a) we show the predictions of the sum rules for
150 realizations of the speckle potential, with $V_s\sim$ 30~Hz.
Each realization is obtained by taking successive 1D slices of
1~$\mu$m from the same 2D experimental speckle pattern. The
average shift is close to zero for both excitation modes. We have
performed a similar series of measurements experimentally.
Fig.~\ref{fig4}(b) shows the quadrupole frequency for different
realizations of the speckle potential at $V_s\sim$80~Hz. As
predicted, we observe both red and blue shifts, however within our
limited statistics we see a bias towards the red. We have also
begun an exploration of the dipole and quadrupole modes for larger
amplitude oscillations where the solution of the GPE indicates
that the superfluidity of the system may be compromised. The
simulations show short wavelength modulation of the density
distribution and damped oscillations. Also in the experiment we
found first evidence of increased quadrupole frequency shifts and
strong perturbations of the expanded density profile in this
regime.

In summary, we present the first results on a BEC in a
controllable random speckle potential. We see the effect of a weak
random potential indicated by stripes in the expanded density
profile of the BEC, damped dipole oscillations, and frequency
shifts in the quadrupole mode that are not correlated to the
measured dipole frequency. Simulations solving the GPE and an
analytical sum-rules approach successfully describe these shifts
in the pertubative regime. In the limit of a strong random
potential, we observe a broad gaussian density profile after
expansion, indicating that the atoms remain localized in the
individual wells of the speckle potential. This opens the way to
new investigations of localization phenomena in quantum gases, and
the superfluid behavior with varying roughness of a surface
potential.

This work has been supported by the EU Contracts No.
HPRN-CT-2000-00125, INFM PRA ``Photon Matter'' and MIUR FIRB 2001.
J.~E.~L. was supported by EU. We acknowledge for stimulating
discussions G.~Shlyapnikov, R.~Roth, J.~Zakrzewski, S.~Stringari
and colleagues from the INFM-BEC Center in Trento, and all the
people of the Quantum Gases group in Florence. We also thank
F.~S.~Cataliotti, V.~Guarrera and M.~Scheid for experimental
assistance and discussions.


\begin{thebibliography}{99}

\bibitem[*]{mail}
Electronic address: \verb"lye@lens.unifi.it"

\bibitem{review}
J.~D.~Reppy, J. Low Temp. Phys. \textbf{87}, 205 (1992) and
references therein.

\bibitem{lesssuperfluid} G.~E.~Astrakharchik, J.~Boronat, J.~Casulleras, and S.~Giorgini,
Phys. Rev. A \textbf{66}, 023603 (2002); S.~Giorgini,
L.~Pitaevskii, S.~Stringari, Phys. Rev. B \textbf{49}, 12938
(1994); K.~Huang and H.~F.~Meng, Phys. Rev. Lett. \textbf{69}, 644
(1992); A.~V.~Lopatin and V.~M.~Vinokur, Phys. Rev. Lett.
\textbf{88}, 235503 (2002).



\bibitem{boseglass}
M.~P.~A.~Fisher, P.~B.~Weichman, G.~Grinstein, D.~S.~Fisher, Phys.
Rev. B \textbf{40}, 546 (1989); R.~T.~Scalettar, G.~G.~Batrouni,
and G.~T.~Zimanyi, Phys. Rev. Lett. \textbf{66}, 3144 (1991);
W.~Krauth, N.~Trivedi, and D.~Ceperley, Phys. Rev. Lett.
\textbf{67}, 2307 (1991).

\bibitem{fromdamski}
N.~Markovic, C.~Christiansen, A.~M.~Mack, W.~H.~Huber, and
A.~M.~Goldman, Phys. Rev. B \textbf{60}, 4320 (1999);
P.~A.~Crowell, F.~W.~Van Keuls, and J.~D.~Reppy, Phys. Rev. B 55,
12620 (1997); H.~S.~J.~van der Zant, W.~J.~Elion, L.~J.~Geerligs,
and J.~E.~Mooij, Phys. Rev. B \textbf{54}, 10081 (1996).

\bibitem{anderson58}
P.~W.~Anderson, Phys. Rev. \textbf{109}, 1492 (1958).

\bibitem{soliddefects}
See, e.g., P.~A.~Lee and T.~V.~Ramakrishnan, Rev. Mod. Phys.
\textbf{57}, 287 (1985); Ping Sheng, \textit{Introduction to Wave
Scattering, Localization, and Mesoscopic Phenomena} (Academic
Press, New York, 1995).

\bibitem{Diederik97}
R.~Dalichaouch \textit{et al.}, Nature \textbf{354}, 53 (1991);
D.~S.~Wiersma \textit{et al.}, Nature \textbf{390}, 671 (1997).

\bibitem{latticereview}
J.~H.~Denschlag \textit{et al.}, Journ. of Phys. B \textbf{35},
3095 (2002).

\bibitem{greiner}
M.~Greiner \textit{et al.}, Nature \textbf{415}, 39 (2002).

\bibitem{RothandBurnett}
R.~Roth and K.~Burnett, Phys. Rev. A \textbf{68}, 023604 (2003).

\bibitem{Damski}
B.~Damski, J.~Zakrzewski, L.~Santos, P.~Zoller, and M.~Lewenstein,
Phys. Rev. Lett. \textbf{91}, 080403 (2003).

\bibitem{fragmentation}
S.~Kraft \textit{et al.}, J. Phys. B \textbf{35}, L469 (2002);
A.~E.~Leanhardt \textit{et al.}, Phys. Rev. Lett. \textbf{90},
100404 (2003); J.~Est\`{e}ve \textit{et al.}, Phys. Rev. A
\textbf{70}, 043629 (2004); D.~W.~Wang, M.~D.~Lukin, and
E.~Demler, Phys. Rev. Lett. \textbf{92}, 076802 (2004).

\bibitem{Fortagh}
J.~Fort\'{a}gh, H.~Ott, S.~Kraft, A.~G\"{u}nther, and
C.~Zimmermann, Phys. Rev. A \textbf{66}, 041604(R) (2002).

\bibitem{shlyapnikov} G. Shlyapnikov, private communication.

\bibitem{Sengstock}
S.~Dettmer \textit{et al.}, Phys. Rev. Lett. \textbf{87}, 160406
(2001).

\bibitem{Richard03}
S.~Richard \textit{et al.}, Phys. Rev. Lett. \textbf{91}, 010405
(2003)

\bibitem{DI}
L.~Fallani \textit{et al.}, Phys. Rev. Lett. \textbf{93}, 140406
(2004).

\bibitem{Dalibard}
Z.~Hadzibabic, S.~Stock, B.~Battelier, V.~Bretin, and J.~Dalibard,
Phys. Rev. Lett. \textbf{93}, 180403 (2004).

\bibitem{stringarireview}
F.~Dalfovo \textit{et al.}, Rev. Mod. Phys. \textbf{71}, 463
(1999).

\bibitem{chiara}
C.~Fort \textit{et al.}, Phys. Rev. Lett. \textbf{90}, 140405
(2003).

\bibitem{stringari96}
S.~Stringari, Phys. Rev. Lett. \textbf{77}, 2360 (1996).

\bibitem{griffen97}
A.~Griffin, Wen-Chin Wu, S.~Stringari, Phys. Rev. Lett.
\textbf{78}, 1838 (1997).

\bibitem{Stamper-Kurn98}
D.~M.~Stamper-Kurn, H.-J.~Miesner, S.~Inouye, M.~R.~Andrews, and
W.~Ketterle, Phys. Rev. Lett. \textbf{81}, 500 (1998).

\bibitem{kimura}
T.~Kimura, H.~Saito, and M.~Ueda, J. Phys. Soc. Jpn., \textbf{68},
1477 (1999).

\end{thebibliography}
\end{document}